\documentclass{PoS}

\title{Cluster Algorithm Renormalization Group Method}

\ShortTitle{CARG Method}

\author{\speaker{Guillermo Palma} and David Zambrano
        \\Departamento de Física\\Universidad de Santiago de Chile\\Chile.\\
        E-mail: \email{guillermo.palma@usach.cl}
        }

\abstract{We present a self consistent method based on cluster
algorithms and Renormalization Group on the lattice to study
critical systems numerically. We illustrate it by means of the 2D
Ising model. We compute the critical exponents $\nu$ and $\eta$ and
the renormalization group flow of the probability density function
of the magnetization. The results, compared to the standard Monte
Carlo Renormalization Group proposed by Swendsen \cite{MCRG}, are
very accurate and the method works faster by a factor which grows
monotonically with the lattice size. This allows to simulate larger
lattices in reachable computational times.}

\FullConference{The XXVI International Symposium on Lattice Field Theory\\
         July 14-19 2008\\
         Williamsburg, Virginia, USA}

\begin{document}

\section{The Method}

We consider a general spin model defined in a square lattice of
lattice spacing $a$, and linear size $L$, with periodic boundary
conditions which is defined by the Hamiltonian

\begin{equation}
H=\sum_\alpha K_\alpha S_\alpha
\label{def}
\end{equation}
where each $S_\alpha$ is a combination of the spin variables, for
example $S_1$ = $\sigma_i\sigma_j$ for first neighbors,
$S_2$=$\sigma_l\sigma_m$ for second neighbors, and so on.

In this article we illustrate the method by considering the 2D Ising
model, including up to three even interactions (nearest-neighbor,
second-neighbor, and four spin) and one odd interaction (a weak
magnetic field).

The RG theory allows the computation of the critical properties of a
model. The critical exponents for example, can be obtained from the
linearized RG transformation matrix $T_{\alpha\beta}{}^\ast$, by
computing its eigenvalues. $T_{\alpha\beta}{}^\ast$, defined by

\begin{equation}
T_{\alpha\beta}{}^\ast=\left[~\frac{\partial
K_\alpha{}^{(n)}}{\partial K_\beta{}^{(n-1)}}~\right]_{H^\ast}
\label{LRGT}
\end{equation}
can be obtained numerically from the coupled equations,
\begin{equation}
\frac{\partial\langle~S_\gamma{}^{(n)}~\rangle}{\partial
K_\beta{}^{(n-1)}}= \sum_\alpha\frac{\partial
K_\alpha{}^{(n)}}{\partial
K_\beta{}^{(n-1)}}\frac{\partial\langle~S_\gamma{}^{(n)}~\rangle}{\partial
K_\alpha{}^{(n)}}
\label{RGT}
\end{equation}
where,
\begin{equation}
\frac{\partial\langle~S_\gamma{}^{(n)}~\rangle}{\partial
K_\beta{}^{(n-1)}}=-\langle~S_\gamma{}^{(n)}~S_\beta{}^{(n-1)}~\rangle+\langle~S_\gamma{}^{(n)}~\rangle~\langle~S_\beta{}^{(n-1)}~\rangle
\label{RGT1}
\end{equation}
and
\begin{equation}
\frac{\partial\langle~S_\gamma{}^{(n)}~\rangle}{\partial
K_\alpha{}^{(n)}}=-\langle~S_\gamma{}^{(n)}~S_\alpha{}^{(n)}~\rangle+\langle~S_\gamma{}^{(n)}~\rangle~\langle~S_\alpha{}^{(n)}~\rangle
\label{RGT2}
\end{equation}

The critical exponents are obtained from the eigenvalues of
$T_{\alpha\beta}{}^\ast$ in the standard way. For the Ising model
they are given by the standard relations $\nu=\ln
s/\ln\lambda_1{}^{e}$ and $\eta=d+2-2\ln\lambda_1{}^{o}/\ln b$,
where $\lambda_1{}^{e(o)}$ is the largest even (odd) eigenvalue.

\section{The CARG method}

For the 2D Ising model the Hamiltonian reads
\begin{equation}
H(\sigma )=\frac{J}{k_BT} \sum_{<i,j>} \sigma_i ~ \sigma_j
\end{equation}
where $J>0$ describes a ferromagnetic system. In the thermodynamics
limit this model has a second order phase transition at the
Onsager's critical temperature $T_{c} = 2/ln(1+\sqrt(2))$.

\subsection{Magnetic Susceptibility}

In order to compute the linearized RG transformation matrix
$T_{\alpha\beta}{}^\ast$ from eqn. (\ref{LRGT}), the critical
temperature should be used. One can obtain it by computing the
susceptibility for different lattice sizes as it is shown in figure
1. Its maximum for each lattice size $L$ defines the lattice shifted
critical temperature $T_c(L)$.

\begin{figure}[th]
\centering
\includegraphics[width=.6\textwidth]{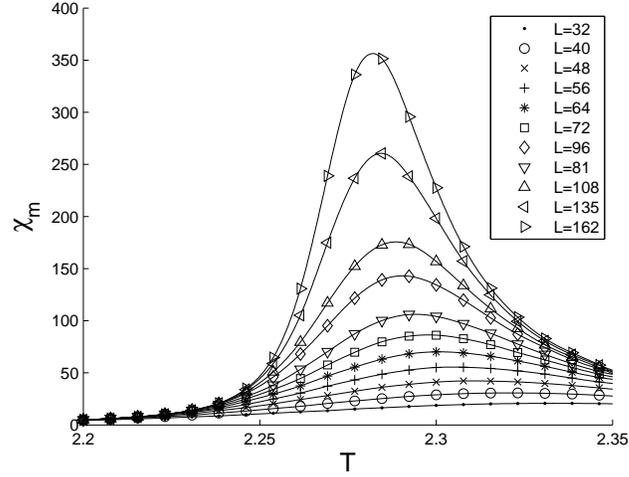}
\caption{The maximum of the susceptibility $\chi_m[L]$ is obtained
using interpolation. For example, $max(\chi_m)[L=108] = 2.2886$,
$max(\chi_m)[L=64] = 2.3008$, and $max(\chi_m)[L=45] = 2.314$.}
\label{fig1}
\end{figure}

\subsection{Finite Size Scaling}

\begin{figure}[th]
\centering
\includegraphics[width=.5\textwidth]{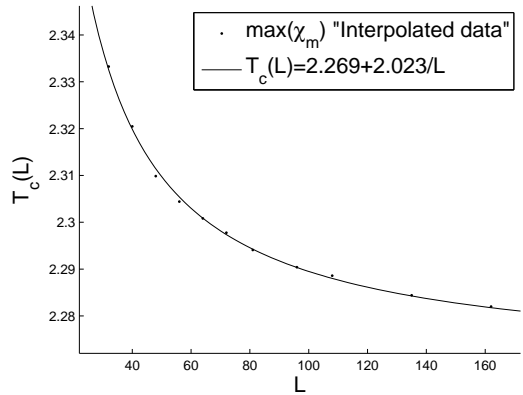}
\caption{From the fit we obtain the numerical values $T_c(\infty) =
2.269$, and $\alpha=2.023$. The convergence to $T_c(\infty)$ is
fairly fast, as can be seen from the following values for the
lattice shifted critical temperature: $T_c(L= 108) = 2.2877$, $T_c(L
= 64)  = 2.3006$, $T_c(L = 45) =2.3140$.}
\label{fig2}
\end{figure}

Now we perform a finite size scaling analysis (FSS) of the numerical
results obtained above to extract the infinite volume limit of the
critical temperature. We use the standard analytical formula
$T_c(L)=T_c(\infty)+\alpha/L$ to fit the numerical values $T_c(L)$.
The fit is shown in figure \ref{fig2}.

\newpage

The thermodynamic limit of the lattice shifted critical temperature
$T_c(\infty)$ agrees with the Onsager's critical value up to an
error lest than 0.8 per one thousand. ~At this point we want to
emphasize that a precise value for exact value of the critical
temperature is no necessary to compute the eigenvalues of the
linearized RG transformation matrix. In fact, it is only necessary
to calculate the derivatives appearing in eqn (\ref{LRGT}) in a
neighborhood of a "linear region", where they are essentially
constant.

\subsection{Binder Cumulant}

\begin{figure}[th]
\centering
\includegraphics[width=.6\textwidth]{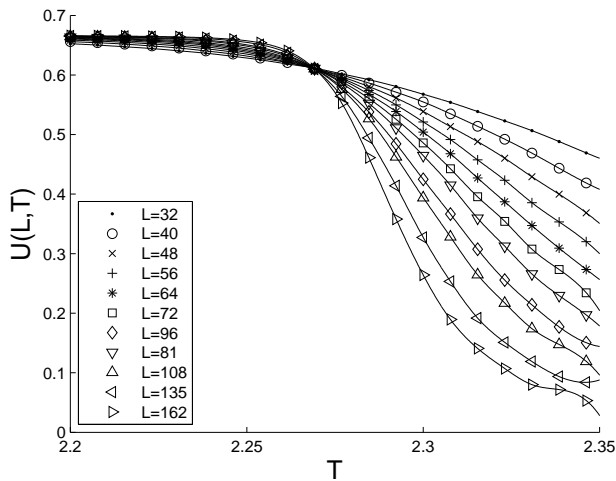}
\caption{The intersection of these curves occurs at: $T_c(\infty) =
2.26903 \pm 0.00059$. This value differs in $1$ percent from
Onsager's exact result.}
\label{fig3}
\end{figure}

One alternative method to compute the critical temperature of the
infinite volume system is to use the Binder Cumulant

\begin{equation}
u(L,T)=1-\frac{1}{3}\frac{\langle~M{}^{4}~\rangle}{\langle~M{}^{2}~\rangle
^{2}}. \label{BC}
\end{equation}
which becomes independent of the lattice size at the Onsager's
critical temperature. This behavior is shown in figure \ref{fig3}.

The result obtained with the Binder Cumulant is consistent with the
above result obtained by a FSS analysis.

\section{Numerical Results}

\subsection{Computing the critical exponents.}

To perform the simulation on the fundamental lattice we used the
Wolff algorithm \cite{UW}. A total of $10^5$ sweeps were used to
thermalize the system and $10^6$ configurations were used to compute
thermal averages. Up to three even interactions (nearest-neighbor,
second-neighbor and four spin or plaquette) and one odd interaction
(magnetic field) were considered. We considered a lattice of lattice
size $L=64$ and performed two RG steps and used the renormalization
parameter $b=2$. The results for the critical exponents $\nu$ and
$\eta$ are displayed in the following table. Compared to the exact
values, the numerical results agree remarkably well, with errors of
one in one thousand.

\begin{table}[th]
\begin{center}
\begin{tabular}{|c||c|c|c|c|}
\hline
\hline RG Step & $\lambda_1{}^e$ & $\lambda_1{}^o$ & $\nu$    & $\eta$   \\
\hline $1$     & $1.9586$        & $3.6856$        & $1.0311$ & $0.2362$ \\
\hline $2$     & $1.9986$        & $3.6676$        & $1.0010$ & $0.2503$ \\
\hline Exact   & $2$             & $3.6680$        & $1$      & $0.250$  \\
\hline \hline
\end{tabular}
\end{center}
\caption{Critical exponents for $L=64$ and $T=2.259$.}
\label{tab.1}
\end{table}

In order to compare our results with the ones obtained in reference
\cite{MCRG2}, we considered a lattice of lattice size $L=108$ and
performed three RG steps using the renormalization parameter $b=3$.
The results are shown in the following table, where the label [S]
stands for the result of ref. \cite{MCRG2} and [PZ] for ours.

\begin{table}[th]
\begin{center}
\begin{tabular}{|c||c|c|c|c|}
\hline
\hline RG Step     & $\lambda_1{}^e$ & $\lambda_1{}^o$ & $\nu$     & $\eta$   \\
\hline $1_{[S]}$   & $2.852$         & $7.705$         & $1.048$   & $0.2828$ \\
\hline $1_{[PZ]}$  & $2.8635$        & $7.7062$        & $1.0443$  & $0.2825$ \\
\hline $2_{[S]}$   & $3.021$         & $7.828$         & $0.994$   & $0.2540$ \\
\hline $2_{[PZ]}$  & $3.0027$        & $7.8269$        & $0.9992$  & $0.2542$ \\
\hline $3_{[S]}$   & $3.007$         & $7.831$         & $0.998$   & $0.2534$ \\
\hline $3_{[PZ]}$  & $3.0013$        & $7.8361$        & $0.9996$  & $0.2521$ \\
\hline Exact       & $3$             & $7.8452$        & $1$       & $0.250$  \\
\hline \hline
\end{tabular}
\end{center}
\caption{The critical exponents obtained by the CARG method [PZ] and
by Swendsen's method [S] for $L=108$ are displayed.}
\label{tab.2}
\end{table}

The numerical results for the critical exponents $\nu$ and $\eta$
are very accurate and both results ([S] and [PZ]) agree remarkable
well with their corresponding exact values. The main advantage of
our method consists that it is faster than the method proposed by
\cite{MCRG2} by a factor which increases with the system size.

\section{RG-flow of the PDF}

\subsection{Probability Density Function}

As it was pointed out in \cite{MPV}, the probability density
function (PDF) of the magnetization corresponds to the partition
function of an auxiliary theory, which equals the original theory
plus a very small imaginary coupling. From RG theory one knows that
partition functions are invariant under RG transformations. This is
confirmed directly in the following figure.

\begin{figure}[th]
\centering
\includegraphics[width=.6\textwidth]{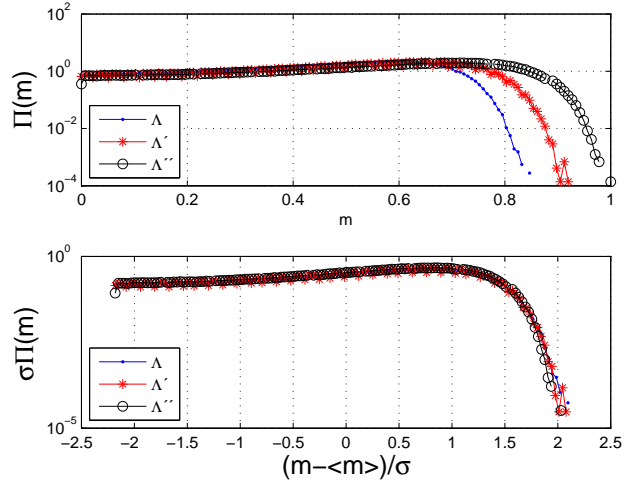}
\caption{$L = 64$ at $T = T_c(L) = 2.3008$. $\Lambda ~~\rightarrow L
= 64$. $\Lambda' ~\rightarrow L = 32$. $\Lambda'' \rightarrow L =
16$.}
\label{fig4}
\end{figure}

The first plot of figure \ref{fig4} shows the PDF for the
fundamental lattice and for two coarse grained lattices. In the
second plot the PDFs normalized to the first two moments are
displayed. The three curves collapse onto one curve with high
accuracy, showing the scale invariance of the PDF.

\section{Conclusions}

We have shown how to obtain very accurate values for the critical
exponents and how to compute the RG flow of the PDF of the
magnetization without the previous knowledge of the critical
temperature. The method is self consistent and when compared to the
original method proposed in \cite{MCRG2}, our method is faster by a
factor which grows linearly with the lattice size from 6 for $L=64$
until a factor 10 for $L=162$.  The use of a cluster algorithm to
simulate the fundamental Hamiltonian leads to this advantage, which
allows to simulate larger lattices in reasonable computation times.
A further application of the method in connection with universal
fluctuations is reported in \cite{CARGUF}.

\acknowledgments This work was partially supported by DICYT of
University of Santiago de Chile.

\end{document}